\documentclass[doublecol]{epl2}

\renewcommand{\epsilon}{\varepsilon}
\newcommand{\figurewidth}{0.46\textwidth}
\newcommand{\narrowfigurewidth}{0.26\textwidth}

\title{Driven polymer translocation through nanopores: slow versus fast
dynamics}
\shorttitle{Driven polymer translocation through nanopores}

\author{Kaifu Luo\inst{1,2}\footnote{e-mail: luokaifu@gmail.com} \and Tapio Ala-Nissila\inst{3,4}
\and See-Chen Ying\inst{4} \and Ralf Metzler\inst{1}}
\shortauthor{K. Luo \etal}

\institute{
  \inst{1} Physics Department, Technical University of Munich, D-85748 Garching, Germany\\
  \inst{2} Department of Polymer Science and Engineering and CAS Key Laboratory of Soft Matter Chemistry, University of Science and Technology of China, Hefei, Anhui 230026, China\\
  \inst{3} Department of Applied Physics and COMP Center of Excellence, Helsinki University of Technology, P.O. Box 1100, FIN-02015 TKK, Espoo, Finland\\
  \inst{4} Department of Physics, Brown University, P.O. Box 1843, Providence, Rhode Island 02912, USA
}

\pacs{87.15.A-}{Theory, modeling, and computer simulation}
\pacs{87.15.H-}{Dynamics of biomolecules }
\pacs{36.20.-r}{Macromolecules and polymer molecules}

\abstract{
We investigate the dynamics of polymer translocation through nanopores under
external driving by 3D Langevin Dynamics simulations, focusing on the scaling of the average translocation time $\tau$ versus the length of the polymer, $\tau\sim N^{\alpha}$. For slow
translocation, i.e., under low driving force and/or high friction, we find
$\alpha \approx 1+\nu \approx 1.588$ where $\nu$ denotes the Flory exponent. In contrast,
$\alpha\approx 1.37$ is observed for fast translocation due to the highly
deformed chain conformation on the trans side, reflecting a pronounced non-equilibrium situation.
The dependence of the translocation time on the driving force is given by
$\tau \sim F^{-1}$ and $\tau \sim
F^{-0.80}$ for slow and fast translocation, respectively. These results clarify the
controversy on the magnitude of the scaling exponent $\alpha$ for driven translocation.}

\begin{document}
\maketitle

\section{Introduction}
The passage of a polymer through a nanopore is essential
for numerous biological processes such as DNA and RNA translocation across
nuclear pores, protein transport through membrane channels, or virus injection
into cells \cite{alberts}. (Bio)polymer translocation is also at the heart of
various potential technology applications, for instance, rapid DNA sequencing,
gene therapy, or controlled drug delivery \cite{Meller03}. A translocating
polymer has to overcome an entropic barrier. In biological cells, biopolymers
are driven through pores by transmembrane potentials, chemical potential
gradients due to binding proteins, active pulling by polymerase, or entropic
pressure (virus ejection) \cite{alberts}. From both the basic physics
as well as a technology design perspective,
an important measure is the scaling of the average translocation
time $\tau$ with the polymer length $N$, $\tau\sim N^{\alpha}$, and the value
of the corresponding scaling exponent $\alpha$.

Most translocation experiments are carried out under an applied electric field
across the pore. We focus here on this particular type of driven translocation.
In experiments with $\alpha$-hemolysin pores of inner diameter 2 nm a linear behavior
of $\tau\sim N$ ($\alpha=1$) was
observed \cite{Kasianowicz96,MellerPNAS}, while an exponent $\alpha=1.27$ was
obtained for double-stranded DNA translocation through a solid-state nanopore
of inner diameter 10 nm \cite{Storm}. Recently the voltage-driven translocation
of individual DNA molecules through solid-state nanopores of diameters
$2.7\ldots 5$ nm revealed that $\tau\sim N^{1.40}$ for DNA molecules in the
range $150\sim 3500$ base pairs \cite{Meller08}.

Inspired by these experiments, a number of recent theories and simulations
on the translocation dynamics have been presented. Standard Kramers
analysis of diffusion across an equilibrium entropic barrier yields $\tau\sim N^2$ for
unbiased translocation and $\tau\sim N$ under external driving (assuming
friction to be $N$-independent) \cite{Sung,Muthukumar}. However, as noted in
Ref.~\cite{Chuang} the quadratic scaling behavior in the unbiased case cannot
be correct for a self-avoiding polymer, as the translocation time would be
shorter than the Rouse equilibration time scaling like $\tau_R\sim N^{1+2\nu}$,
involving the Flory exponent $\nu$ ($\nu=0.588$ in 3D, $\nu=0.75$ in 2D) \cite{gennes}.
This finding renders the concept of equilibrium entropy and the ensuing picture of
entropic barrier crossing inappropriate for translocation dynamics.
Numerically, it was shown that for large $N$, $\tau\sim N^{1+2\nu}$, i.e., the translocation time
scales in the same way as the equilibration time, but with a much larger prefactor
\cite{Chuang}. This result was recently corroborated by extensive numerical
simulations based on the Fluctuating Bond (FB) \cite{Luo06} and Langevin
Dynamics (LD) models with the bead-spring approach
\cite{Huopaniemi06,Milchev,Liao}.
For \emph{driven} translocation, the deviation from equilibrium is expected to be
even more pronounced,
and the equilibrium entropic barrier is less relevant for the translocation dynamics.
For this case,  a lower bound $\tau
\sim N^{1+\nu}$ was estimated  for the translocation time \cite{Kantor}. Lattice Monte Carlo
(MC) simulations of self-avoiding chains in $2D$ revealed $\alpha\approx1.5$
\cite{Kantor}, which is smaller than $1+\nu=1.75$, a difference attributed to
finite size effects. However, additional simulation studies
found a crossover
from $\tau\sim N^{1.46\pm0.01} \approx N^{2\nu}$ for relatively short polymers
to $\tau\sim N^{1.70\pm0.03} \approx N^{1+\nu}$ for longer chains ($N>200$) using FB
\cite{Luo062} and LD \cite{Huopaniemi06,Luo07} models in 2D.
Increasing the friction, the crossover vanished and only exponent $1+\nu$ was
observed \cite{Huopaniemi06}.
However, the crossover is absent in 3D for quite long chains ($N \sim 800$) and $\alpha \approx 1.40$ is observed using both LD \cite{Bhatta08} and GROMACS molecular dynamics with LD thermostats \cite{LuoPRE08} simulations.
Lehtola \textit{et al.} \cite{Linna08} find approximately the same exponent of about 1.40 based on LD simulations, however, their exponent increases with increasing force, which cannot hold asymptotically.
Moreover, Gauthier and Slater \cite{Slater08} reported $\tau \sim N^{1+\nu}$
using an exact numerical method valid at low bias.

Recently, however, alternative scaling scenarios have been presented in Refs.
\cite{Panja,Dubbeldam072} that contradict above results. These
 two views disagree with each other \cite{DubbeldamC}. To resolve the apparent
discrepancy on the value of $\alpha$ for driven translocation, we here report
results on the behavior of $\tau$ as a function of $N$ from high-accuracy LD
simulations in 3D. As in the above-mentioned studies, we also neglect
hydrodynamics effects as well as polymer-pore interactions
\cite{Luo072}. We find that there exists two limiting
 regimes, corresponding to slow and  fast translocation respectively. The
slow translocation case is realized for  low driving forces and/or high friction,
and in this regime $\alpha \approx1+\nu$.
In the opposite limit of  fast translocation for high driving forces and/or
low friction, the corresponding scaling exponent is given by
$\alpha\approx 1.37$. As we will argue below,  the difference between the scaling exponents for
these two regimes can be ascribed to the highly deformed chain conformation during fast
translocation, reflecting a pronounced non-equilibrium situation. Our results
clarify the controversy on the value of  $\alpha$ for driven translocation.

\section{Model and method}
\begin{figure}[tb]
\onefigure[width=\narrowfigurewidth]{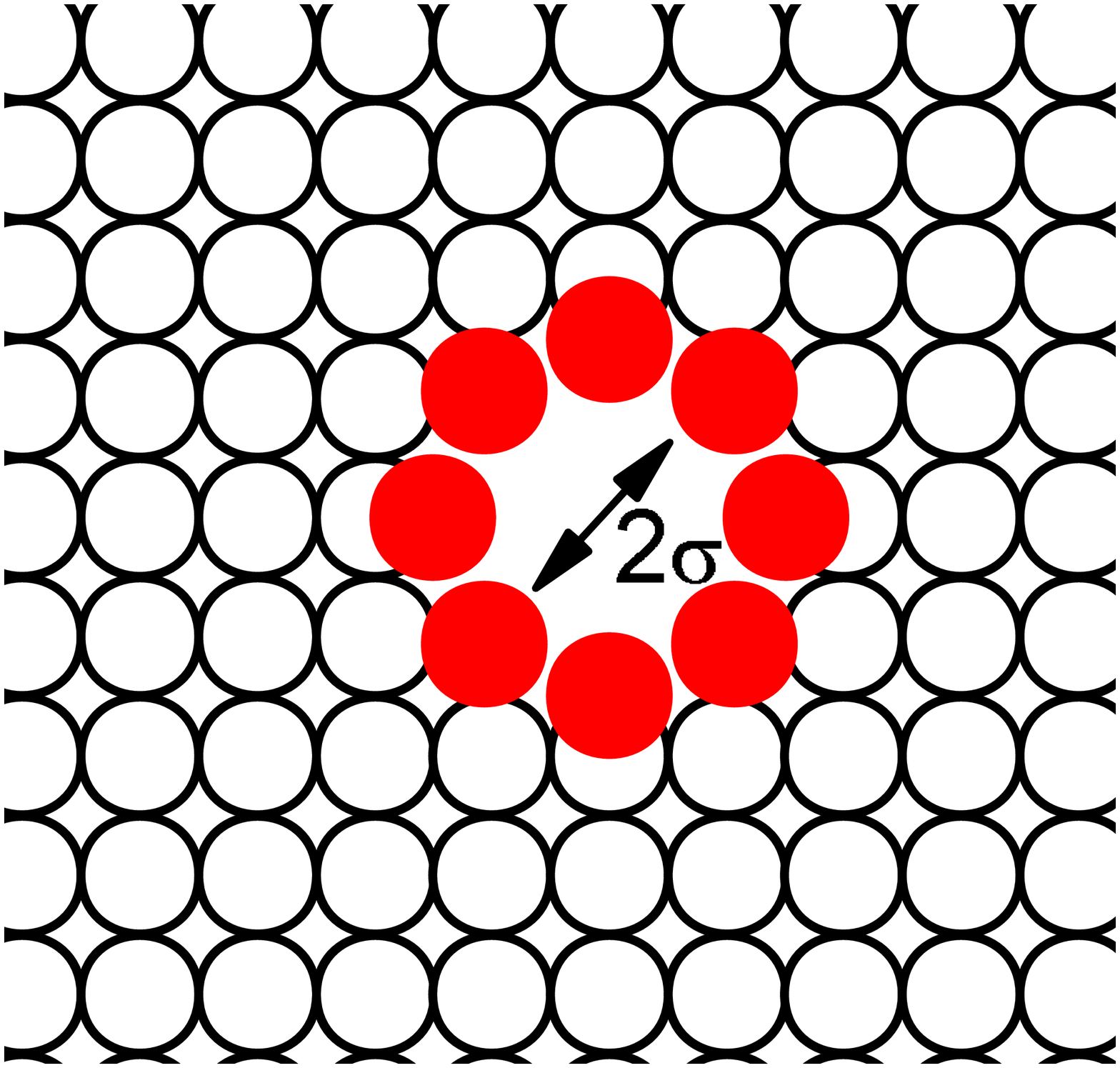}
\caption{Schematic representation of our simulation
system. The wall consists in a single layer of beads while the pore itself is
formed by eight particles with their centers equally
distributed on a circle of diameter $3\sigma$.
Therefore, the actual pore diameter is $2\sigma$.}
\label{Fig1}
\end{figure}

In our simulations, the polymer chains are modeled as
bead-spring chains of Lennard-Jones (LJ) particles with the Finite Extension
Nonlinear Elastic (FENE) potential. Excluded volume interaction between
monomers is modeled by a short range repulsive LJ potential: $U_{LJ}(r)=4
\epsilon[{(\frac{\sigma}{r})}^{12}-{(\frac{\sigma} {r})}^6]+\epsilon$ for $r\le
2^{1/6}\sigma$, and 0 for $r>2^{1/6}\sigma$. Here, $\sigma$ is the monomer
diameter and $\epsilon$ is the potential depth. The connectivity between
neighboring monomers is modeled as a FENE spring with $U_{FENE}(r)=-
\frac{1}{2}kR_0^2\ln(1-r^2/R_0^2)$, where $r$ is the distance between
consecutive monomers, $k$ the spring constant, and $R_0$ the maximum allowed
separation between connected monomers. The wall (``membrane'') carrying the
pore is composed of particles of diameter $\sigma$. The wall thickness is
$\sigma$. The nanopore consists of eight particles with their centers equally
distributed on a circle of diameter $3\sigma$, see Fig.~\ref{Fig1}.
Therefore, the actual pore diameter is $2\sigma$. Between all monomer-wall particle
pairs, there exists the same short range repulsive LJ interaction as described above.

In LD simulations each monomer is subjected to conservative, frictional, and
random forces:
$m{\bf\ddot{r}}_i=-{\bf\nabla}({U}_{LJ}+{U}_{FENE})-\xi\dot{\bf
r}_i+{\bf F}_{\textrm{ext}}+{\bf F}_i^R$
\cite{Allen}, where $m$ is the monomer mass, $\xi$ the
friction coefficient, and ${\bf F}_i^R$ the random force, that satisfies the
fluctuation-dissipation theorem. The external force is expressed as
${\bf F}_{\textrm{ext}}=F\hat{x}$, where $F$ is the external force strength
exerted on the monomers in the pore, and $\hat{x}$ is a unit vector in the
direction along the pore axis.
In the present work, we use the LJ parameters
$\epsilon$ and $\sigma$, and the monomer mass $m$ to fix the energy, length and
mass scales. This sets the time scale $t_{LJ}=(m\sigma^2/\epsilon)^{1/2}$.
The dimensionless parameters in our simulations are $R_0=2$, $k=7$, $k_{B}T=
1.2$ and $\xi=0.7$, unless otherwise stated. The Langevin equation is
integrated in 3D by a method described in Ref.~\cite{Ermak}.
Initially, the first monomer of the chain is placed in the entrance of the
pore, while the remaining monomers evolve in the Langevin thermostat to obtain
an equilibrium configuration. The translocation time is defined as the time
interval between the entrance of the first segment into the pore and the exit
of the last segment. Typically, we average our data over 1000 independent runs.

We note that in our model hydrodynamic interactions and the
polymer-pore interactions are neglected. Regarding the issue of hydrodynamics,
recent Molecular Dynamics~\cite{Slater} and Lattice Boltzmann~\cite{Pablo,Fyta}
results show that hydrodynamics is screened out in a narrow pore, which is the
case studied here as well as in the experiments.
In Ref. \cite{Linna09}, however, minor increases of scaling exponents with
increasing driving force are obtained in the presence of hydrodynamic interactions.
Recently, we have used LD simulations
to investigate the influence of an attractive component of polymer-pore interactions on
the  translocation dynamics~\cite{Luo072}.
We found that with increasing strength of the attractive interaction, the histogram for
the translocation time $\tau$ shows a transition from a Gaussian distribution to a long-tailed
distribution corresponding to thermal activation over a free energy barrier.
In addition, a strong attractive polymer-pore interaction can directly affect the
effective scaling exponents of $\tau$ both with $N$ and with the applied voltage, which
provides a possible explanation for the different experimental findings on these physical quantities.
However, our main focus here is to arrive at a detailed understanding of the translocation
dynamics for a purely repulsive pore, and to compare with the results in Refs. \cite{Panja,Dubbeldam072}
to settle the controversies regarding the
scaling behavior of translocation time versus the length of the polymer. Hence, the attractive
part of the polymer-pore interaction is left out in this study.

\section{Results and discussion}
As defined above, our pore length is $\sigma$ and the pore diameter is $2\sigma$.
Previous studies \cite{Luo062} showed that for
longer pores there is no obvious power-law scaling for relatively short chains.
This may be the reason for a linear dependence reported in Ref. \cite{Tian}
where the pore length is $5\sigma$ with chain length $N<80$ and in
Ref. \cite{Paun} where the pore length is $12\sigma$ with chain length $N<100$.
Various heuristic scaling arguments for $\tau$ have been presented for short pores, e.g., in
Refs.~\cite{Kantor,Grosberg,Luo062,Dubbeldam072,Panja} and will be compared to
our numerical results below.

\begin{figure}[tb]
\onefigure[width=\figurewidth]{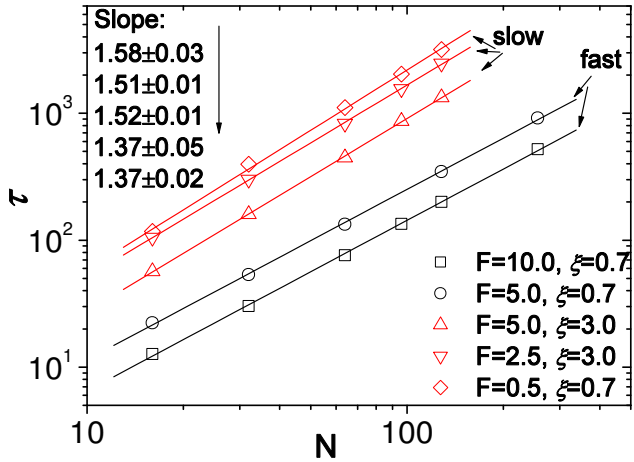}
\caption{Translocation time $\tau$ versus chain length $N$ for different
driving forces $F$ and friction coefficients $\xi$. For fast translocation
(low driving force or high friction) $\alpha \approx 1.37$; for slow
translocation $\alpha\approx1+\nu$.}
\label{Fig2}
\end{figure}

\begin{figure}[tb]
\onefigure[width=\figurewidth]{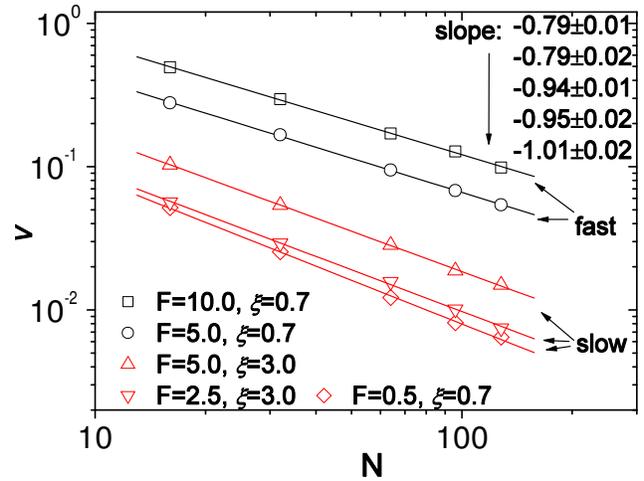}
\caption{Scaling of translocation velocity with chain length.}
\label{Fig3}
\end{figure}

The translocation time as a function of the polymer length is plotted in
Fig.~\ref{Fig2}. One of the main features is that $\alpha$ depends
significantly on both driving force and friction.
For stronger driving forces ($F=10.0$ and 5.0) and lower friction ($\xi=0.7$),
we find $\alpha={1.37\pm0.02}$ and $\alpha={1.37\pm0.05}$,
which are in agreement with previous results \cite{LuoPRE08,Bhatta08,Sakaue} and the
recent experimental findings for individual short DNA molecules in the range
$150\sim3500$ bps through solid-state nanopores \cite{Meller08}.
However, decreasing $F$ or increasing $\xi$, we observe
$\alpha={1.58\pm0.03}$, ${1.51\pm0.01}$ and ${1.52\pm0.01}$ for $F=0.5$ and $\xi=0.7$, $F=2.5$
and $\xi=3.0$, and $F=5.0$ and $\xi=3.0$, respectively.
These exponents are in good agreement, not only with the prediction $1+\nu$ from
Ref.~\cite{Kantor,Grosberg}, but also with the results from the exact numerical
method reported in Ref.~\cite{Slater08} for low fields, as well as with our previous 2D
simulations for relatively long polymers \cite{Luo062,Huopaniemi06,Luo07}.
These observations demonstrate that there exist two limiting dynamic
regimes as demonstrated in Fig.~\ref{Fig2}: slow and fast translocation.
For slow translocation, $\alpha\approx1+\nu$, while $\alpha\approx1.37$ for fast
translocation.

In the scaling arguments presented in Ref.~\cite{Kantor}, an essential
assumption is that the chain is not severely deformed during translocation,
and in particular the chain configuration of the already translocated
part of the chain is close to equilibrium. For slow translocation, these assumptions are
satisfied.
Thus, it is not surprising that the predicted exponent $1+\nu$ is observed in
this regime, in contrast to the fast dynamics regime claimed in
Ref.~\cite{Linna08}. Conversely, for fast translocation the chain is highly deformed
and the translocation dynamics is different. In our previous 2D simulations
based on FB \cite{Luo062} and LD \cite{Huopaniemi06,Luo07}, a crossover from
$\tau\sim N^{2\nu}$ for relatively short polymers ($N<200$) to $\tau\sim
N^{1+\nu}$ for longer chains was found under a stronger driving force $F=5$ and
friction $\xi=0.7$. Increasing $N$ also slows down the translocation dynamics
and thus the system changes from the fast to the slow translocation regime.
However, for the same $F$ and $\xi$ values we failed to observe a similar crossover
for $N=40\sim 800$ in 3D \cite{LuoPRE08}, possibly due to the fact  that the
value of $N$ for the crossover is at a much higher value as compared to the 2D
situation. To actually observe the crossover, an alternative choice is to lower the driving force
or increase the
friction.

Based on the fractional Fokker-Planck equation involving long-range memory
effects (compare Ref.~\cite{report}), the scaling $\alpha=2\nu +1-\gamma_1$ was
obtained, such that $\alpha=1.55$ in 2D and $1.5$ in 3D; here $\gamma_1$ ($\approx 0.945$
2D and $\approx 0.68$ in 3D) is the critical surface exponent \cite{Dubbeldam072}. The MC
simulations in Ref.~\cite{Dubbeldam072} were carried out for weak driving
forces, producing $\tau\sim N^{1.5}/F$ in 3D, in excellent agreement with the
value $\tau\sim N^{1+\nu}/F$ for slow translocation processes.
Moreover, using linear response theory with memory effects \cite{DubbeldamC},
Vocks {\em et. al.} \cite{Panja} came up with an alternative estimate $\tau \sim
N^{\frac{1+2\nu}{1+\nu}}$ for $3D$, which means $\alpha=1.37$ in $3D$. Their
$\alpha$ in 3D is consistent with our numerical data for fast translocation.
However, their estimate fails to capture the scaling exponent for slow
translocation.

For the driven translocation experiments with a voltage applied across the
pore, the force $F$ acts only on
the few monomers inside the pore. As a consequence, it was argued \cite{Kantor}
that the center of mass of polymer should move with a velocity $v \sim F/N$.
Thus the lower bound for the translocation time of an unhindered polymer is the
time to move through a distance $R_g$ (radius of gyration of the polymer),
giving rise to the scaling $\tau\sim R_g/v\sim N^{1+\nu}/F$.
Fig.~\ref{Fig3} shows the translocation velocity $v$ as function of the
polymer length for different driving forces and friction coefficients, $v\sim N^{\delta}$. Here
$v$ is the average velocity with respect to the last monomer (for details see
Ref.~\cite{Luo062}), and we checked that the corresponding $v$ of the center of
mass scales in the same way.
For slow translocation, we find $\delta=-1.01\pm0.02$ for $F=0.5$ and $\xi=0.7$, $\delta=-0.95\pm0.02$ for $F=2.5$ and
$\xi=3.0$, and $\delta=-0.94\pm0.01$ for $F=5.0$ and $\xi=3.0$,
which are in good agreement with $v\sim N^{-1}$.
For fast translocation the velocity decreases less rapidly with $N$: $\delta=-0.79\pm0.01$ for $F=10.0$ and $\xi=0.7$ and $\delta=-0.79\pm0.02$ for $F=5.0$ and $\xi=0.7$, as observed in
Ref.~\cite{Bhatta08}.
Based on the values of $F/\xi$ we roughly distinguish the slow and fast
regimes, \revision{see Table 1.}
For fast translocation with $\xi=0.7$, $F/\xi=14.28$ for $F=10.0$ and $F/\xi=7.14$
for $F=5.0$, which are much higher than those for the slow translocation where $F/\xi=0.71$ for
$F=0.5$ and $\xi=0.7$, $F/\xi=0.83$ for $F=2.5$ and $\xi=3.0$, and $F/\xi=1.67$ for $F=5.0$ and $\xi=3.0$.

\begin{table*}
\caption{\revision{Summary of numerical results. Here, $F$ is the driving force, $\xi$ the friction coefficient, $\alpha$ the scaling exponent of translocation time $\tau$ as a function of the chain length $N$, $\delta$ the scaling exponent of translocation velocity as a function of $N$, and $\beta$ the scaling exponent of the translocation coordinate $s$ as a function of time.  These results clearly demonstrate two regimes.} }
\begin{center}
\begin{tabular}{llllllll}
\hline

$$ & $F$ & $\xi$ & $F/\xi$ & $\alpha$ ($\tau\sim N^{\alpha}$) & $\delta$ ($v\sim N^{\delta}$) & $\beta$ ($\langle s(t)\rangle\sim t^{\beta}$) & $\alpha\beta$\\
\hline
$$ Fast translocation & 10.0 & 0.7 & 14.28 & $1.37\pm0.02$ & $-0.79\pm0.01$ & $0.84\pm0.01$ & 1.15 \\
$$ & 5.0 & 0.7 & 7.14 & $1.37\pm0.05$ & $-0.79\pm0.02$ & $0.85\pm0.01$ & 1.16 \\
\hline
$$ Slow translocation & 5.0 & 3.0 & 1.67 & $1.52\pm0.01$ & $-0.94\pm0.01$ & $0.71\pm0.01$ & 1.08 \\
$$ & 2.5 & 3.0 & 0.83 & $1.51\pm0.02$ & $-0.95\pm0.02$ & $0.69\pm0.01$ & 1.04 \\
$$ & 0.5 & 0.7 & 0.71 & $1.58\pm0.03$ & $-1.01\pm0.02$ & $0.64\pm0.01$ & 1.01 \\

\hline
\end{tabular}
\end{center}
\end{table*}

\begin{figure}[tb]
\onefigure[width=\figurewidth]{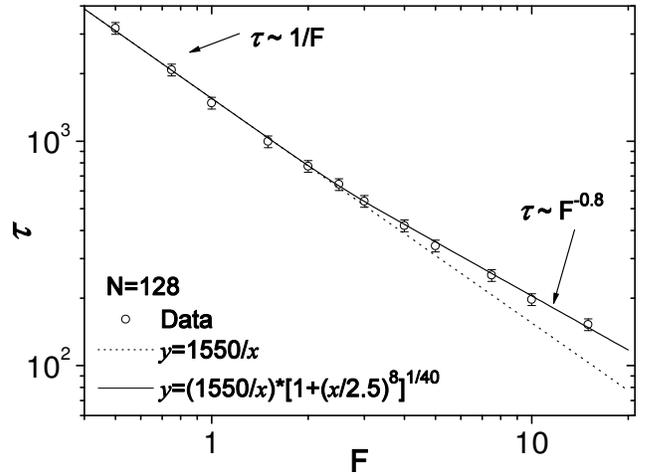}
\caption{Translocation time $\tau$ versus driving force at $\xi=0.7$. The solid line shows the empirical fitting function for the data, as discussed in the text.}
\label{Fig4}
\end{figure}

\begin{figure}[tb]
\onefigure[width=\figurewidth]{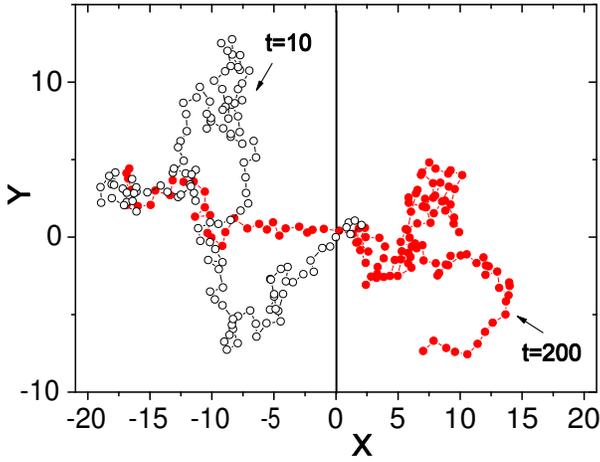}
\caption{Typical chain conformation during fast translocation for $N=128$,
$\xi=0.7$, and $F=5.0$. 3D conformations are projected onto the XY plane.
Black: early stage. Red: Later stage.}
\label{Fig5}
\end{figure}

\begin{figure}[tb]
\onefigure[width=\figurewidth]{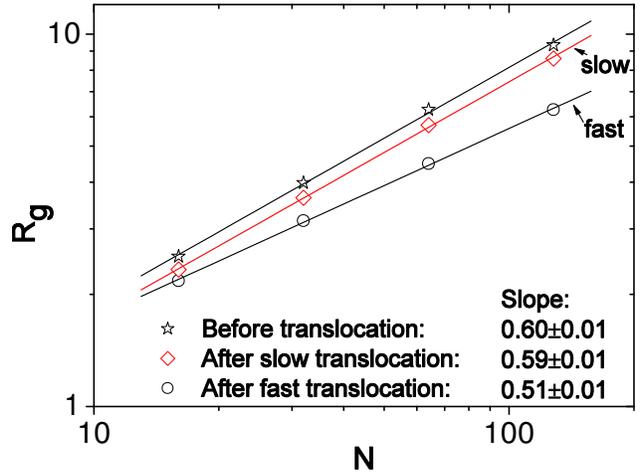}
\caption{The radius of gyration
($R_g$) before translocation and at the moment just after the fast translocation
($\xi=0.7$ and $F=5.0$), and the slow translocation ($\xi=0.7$ and $F=0.5$).}
\label{Fig6}
\end{figure}

Following the change of the scaling exponent $\alpha$, the scaling of the average
velocity changes from $v\sim N^{-0.79}$ for fast translocation,
to $v\sim N^{-1}$ in the case of slow translocation. Furthermore,
we have also studied how the translocation time varies as a function of the driving force for $\xi=
0.7$ and $N=128$ (Fig.~\ref{Fig4}). As long as $F\le2$ we observe $\tau \sim
1/F$, before crossing over to $\tau \sim F^{-0.80}$ for strong driving.
The data can be fitted with an empirical function
$\tau=1550F^{-1}[1+(F/2.5)^{8}]^{1/40}$,
yielding the details of the crossover from the slow translocation (weak force) to the fast
translocation (strong force) regime for this case.
The seemingly high exponent 8 in the empirical function is necessary to account for the
relatively fast turnover between the scaling $\tau \sim F^{-1}$ at low
force and the behavior $\tau \sim F^{-0.8}$ at high force.
The difference of the translocation dynamics in the two regimes can be understood by
inspecting the polymer configurations during the
translocation process. For faster translocation, only  part of the chain on the cis
side can respond immediately, while the remaining part near the chain end does
not feel the force yet. As a result, a part of the chain on the cis side is
deformed to a trumpet and even stem-and-flower shaped \cite{Sakaue}, while the
translocated portion on the trans side has a compact spherical shape, as it does not
have time to diffuse away from the pore exit, see Fig.~\ref{Fig5}. With
increasing time, the tension propagates along the chain, which changes the
chain conformation progressively. Thus, the
chain cannot achieve a steady state with average velocity $v\sim N^{-1}$ even
in the late stage of the translocation process.
Fig.~\ref{Fig6} shows the radius of gyration ($R_g$)
at the moment just after translocation. For fast translocation, this obeys the scaling behavior
$R_g\sim N^{0.51}$,  significantly different
from the equilibrium scaling of $R_g\sim N^{0.6}$,
indicating a non-equilibrium compactification
of the chain immediately after translocation. On the other hand,
the corresponding scaling behavior for the chain after slow
translocation is approximately the same as the chain in equilibrium.

\begin{figure}[tb]
\onefigure[width=\figurewidth]{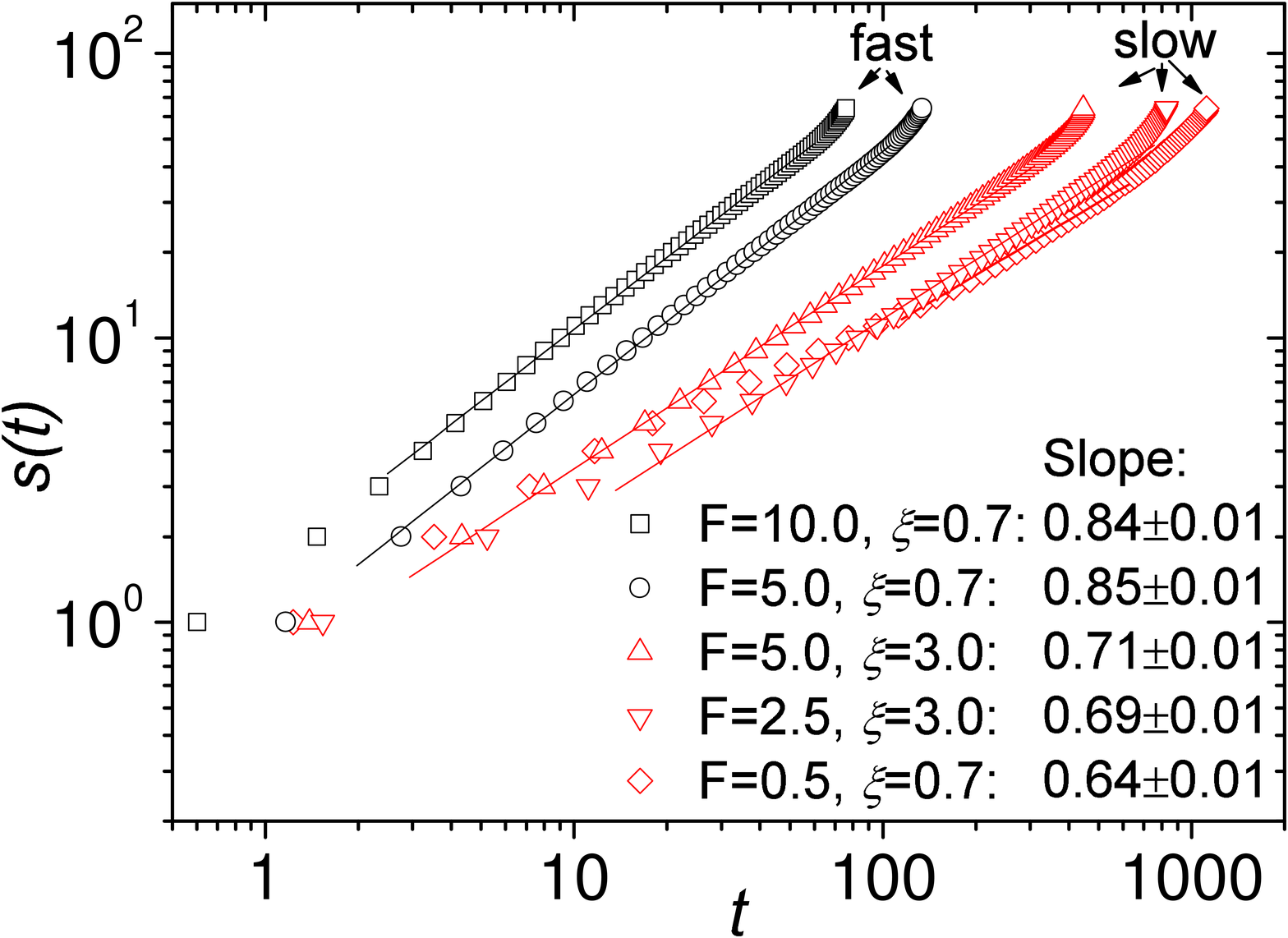}
\caption{The translocation coordinate ($s$-coordinate) as a function of time
for different driving forces and friction coefficients, for $N=64$.}
\label{Fig7}
\end{figure}

We note that in a recent theoretical study the effect of a trumpet shape of
the chain on the cis side, on the translocation dynamics was found to cause a
breakdown of the $\tau\sim 1/F$ scaling \cite{Sakaue}. However, this theory
neglects effects due to the compacted chain structure on the trans side.

We also checked the translocation coordinate $\langle s(t)\rangle\sim t^\beta$
for $N=64$ for different $F$ and $\xi$, see Fig. 7 and \revision{Table 1}.
For fast translocation with $\xi=0.7$, we find $\beta=0.84$ and 0.85 for $F=10.0$ and
5.0, respectively.
However, for slow translocation $\beta=0.71$ for $F=5.0$ and $\xi=3.0$, $\beta=0.69$ for $F=2.5$ and $\xi=3.0$, and $\beta=0.64$ for $F=0.5$ and $\xi=0.7$.
Different $\beta$ values for
fast and slow translocation processes demonstrate the existence of different
dynamic regimes.
The definition $\langle s(t)\rangle\sim t^\beta$ implies that $N \sim \tau^\beta$.
Compared with $\tau \sim N^\alpha$, one obtains $\alpha\beta=1$.
For $F=0.5$ and $\xi=0.7$, $\alpha\beta=1.01 \approx 1$, which
indicates that non-equilibrium effect does not seem to be severe.
However, $\alpha\beta=1.16$ for $F=5.0$ and $\xi=0.7$ indicating a breakdown of
``simple" scaling due to highly non-equilibrium effect.

\section{Conclusions}
We have investigated the dynamics of driven polymer
translocation through nanopores by 3D Langevin dynamics simulations, focusing
on the scaling of the average translocation time $\tau$ as a function of the
polymer length $N$. For
slow translocation, i.e., under low driving forces and/or high friction, we
find $\tau \sim N^{\alpha}$ with $\alpha \approx 1+\nu$. In the opposite case, we obtain
$\alpha \approx 1.37$.
As a function of the driving force $F$, the dependence $\tau \sim 1/F$ and
$\tau \sim F^{-0.80}$ are obtained respectively for slow and fast translocation.
The different behavior in
the two regimes can be understood from analysis of
the chain conformations during the translocation process. In the slow translocation case,
the configurations at all times are close to the equilibrium
case while for the fast translocation regime, there exist highly deformed, unrelaxed chain conformations throughout the translocation process. These
results clarify the controversy on the value of $\alpha$ for driven
translocation in the existing literature.

\acknowledgments
This work has been supported in part by the Deutsche Forschungsgemeinschaft
(DFG). T.A.N. acknowledges the support from The Academy of Finland through its
Center of Excellence (COMP) and TransPoly Consortium grants. We also
acknowledge CSC - the IT Center of Science Ltd. for allocation of computational resources.
K.L. Thanks T. Sakaue for fruitful discussion.

\end{document}